\documentstyle[sprocl,epsfig]{article}

\def\Journal#1#2#3#4{{#1} {\bf #2}, #3 (#4)}

\def\NPB{{\em Nucl. Phys.} B}
\def\PLB{{\em Phys. Lett.}  B}

\def\PRD{{\em Phys. Rev.} D}

\newcommand{\bm}[1]{\mbox{\boldmath $#1$}}
\newcommand{\bt}[1]{{\bm #1}_{_T}}
\newcommand{\Pslash}{\kern 0.2 em P\kern -0.56em \raisebox{0.3ex}{/}}
\newcommand{\pslash}{\kern 0.2 em p\kern -0.4em /}
\newcommand{\kslash}{\kern 0.2 em k\kern -0.45em /}
\newcommand{\Sslash}{\kern 0.2 em S\kern -0.56em \raisebox{0.3ex}{/}}
\newcommand{\dslash}{\kern 0.2 em \partial\kern -0.56em \raisebox{0.3ex}{/}}
\newcommand{\xbj}{x_{_B}}
\newcommand{\zh}{z_h}
\newcommand{\bkt}{\bt{k}}
\newcommand{\bpt}{\bt{p}}

\newcommand{\bSt}{\bt{S}}
\def\be{\begin{equation}}
\def\ee{\end{equation}}
\def\bea{\begin{eqnarray}}
\def\eea{\end{eqnarray}}
\begin{document}

\title{
\begin{flushright}
\small
\vspace*{-0.5 cm}
hep-ph/9707339\\
NIKHEF 97-030
\end{flushright}
SPIN STRUCTURE FUNCTIONS IN LEPTOPRODUCTION\footnote{
Talk at the Conference on Perspectives in Hadronic Physics,
Trieste, 12-16 May 1997.}
}

\author{P.J. MULDERS}

\address{NIKHEF and Free University Amsterdam,\\
P.O. Box 41882, NL-1009 DB Amsterdam, the Netherlands}

\maketitle\abstracts{
We discuss a few examples of structure functions for polarized, semi-inclusive
scattering processes to show the richness of structure. Then we indicate
how polarization and particle production can be used to study the quark
and gluon structure of hadrons going further than the well-known 
parton densities and fragmentation functions. }
  
\section{Structure functions}

We start our discussion with the object of interest for 1-particle
inclusive leptoproduction, the hadronic tensor, given by
\bea
&&2M{\cal W}_{\mu\nu}^{(\ell H)}( q; {P S; P_h S_h} )
=\frac{1}{(2\pi)^4}
\int \frac{d^3 P_X}{(2\pi)^3 2P_X^0}
(2\pi)^4 \delta^4 (q + P - P_X - P_h)
\nonumber
\\
&& \hspace{3.5 cm} \times
\langle {P S} |{J_\mu (0)}|P_X; {P_h S_h} \rangle
\langle P_X; {P_h S_h} |{J_\nu (0)}|{P S} \rangle,
\eea
where $P,\ S$ and $P_h,\ S_h$ are the momenta and spin vectors
of target hadron and produced hadron, $q$ is the (spacelike) 
momentum transfer with $-q^2$ = $Q^2$ sufficiently large. 
\begin{figure}[hb]
\begin{center}
\begin{minipage}{8.5 cm}
\epsfxsize=8.5 cm \epsfbox{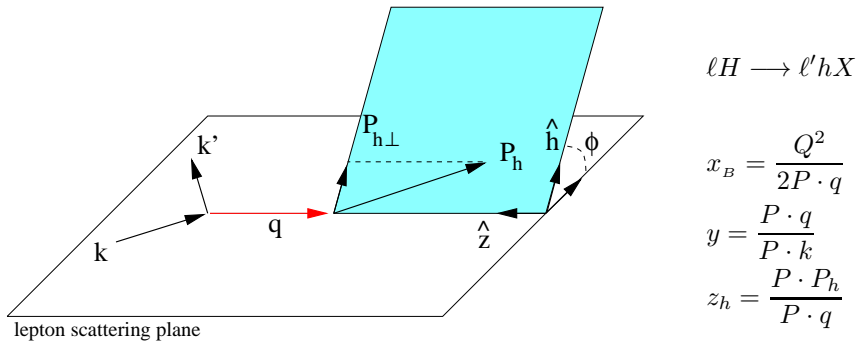}
\end{minipage}
\begin{minipage}{2.5 cm}
\begin{eqnarray*}
&&\ell H \longrightarrow \ell^\prime h X
\end{eqnarray*}
\begin{eqnarray*}
&&\xbj = \frac{Q^2}{2P\cdot q} \\
&&y = \frac{P\cdot q}{P\cdot k}\\
&&z_h = \frac{P\cdot P_h}{P\cdot q}
\end{eqnarray*}
\end{minipage}
\end{center}
\caption{\label{fig1}\em
Kinematics for 1-particle inclusive leptoproduction.}
\end{figure}
The kinematics is illustrated in Fig.~\ref{fig1}, where also the scaling
variables are introduced.
For inclusive scattering (unpolarized lepton and hadron, $\gamma$-exchange)
the most general symmetric part of the hadronic tensor 
is\footnote{
\[
\hat q^\mu = q^\mu/Q, \quad
\hat t^\mu = \tilde P^\mu/\sqrt{\tilde P^2} =
\left(P^\mu - \frac{P\cdot q}{q^2}\,q^\mu\right)/
\sqrt{\tilde P^2}.
\]
}
\be
2MW_S^{\mu\nu}(q,P) =
\underbrace{\left\lgroup - g^{\mu\nu}
+\hat q^\mu \hat q^\nu -\hat t^\mu \hat t^\nu\right\rgroup}_{-g_\perp^{\mu\nu}}
F_1
+ \hat t^\mu\hat t^\nu
\,\underbrace{\left(\frac{F_2}{2\xbj}-F_1\right)}_{F_L}
\ee
Combined with the leptonic part, one obtains the cross section
\be
\frac{d\sigma_O}{d\xbj dy} = \frac{4\pi\,\alpha^2\,\xbj s}{Q^4}
\left\{ \left( 1-y + \frac{1}{2}\,y^2\right) F_T
+ \left( 1 -y\right) F_L\right\}.
\ee
\begin{figure}[t]
\begin{center}
\leavevmode
\epsfxsize=4.5cm \epsfbox{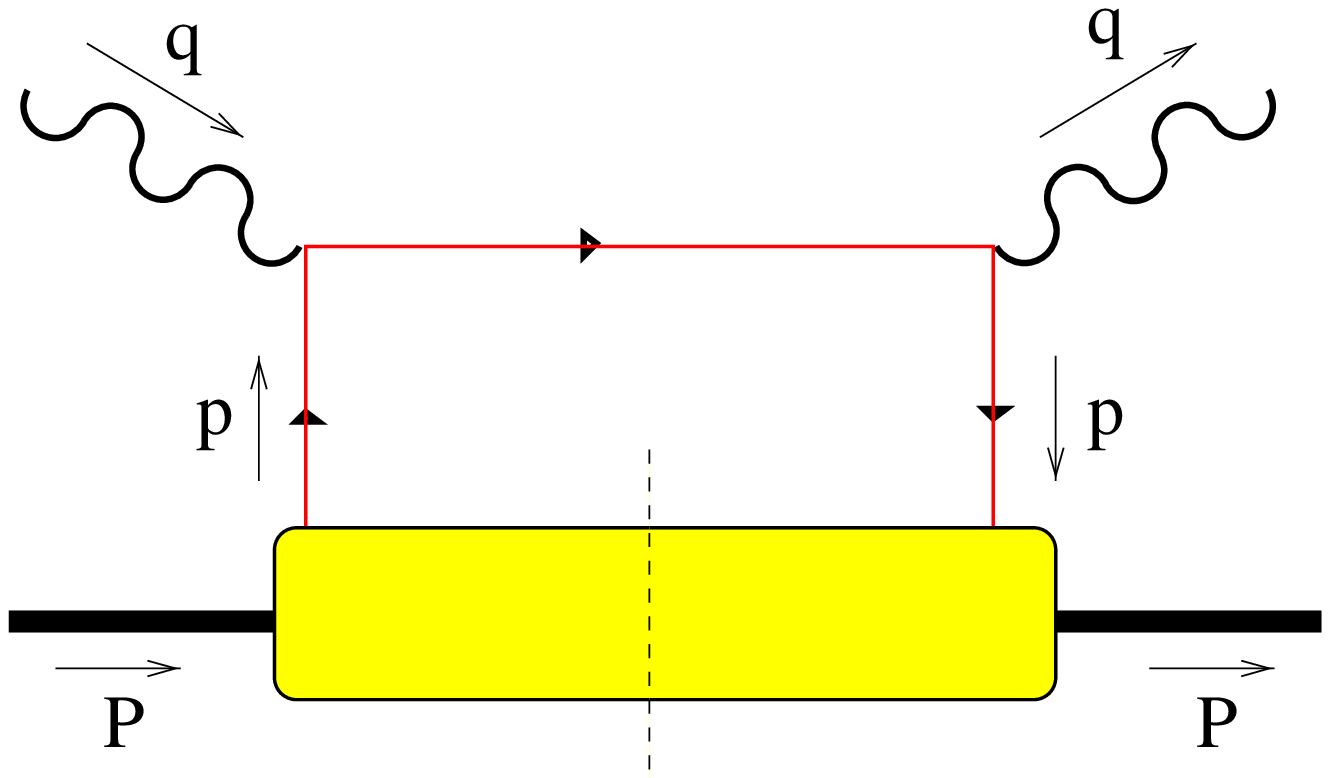}
\hspace{2 cm}
\epsfxsize=4.5cm \epsfbox{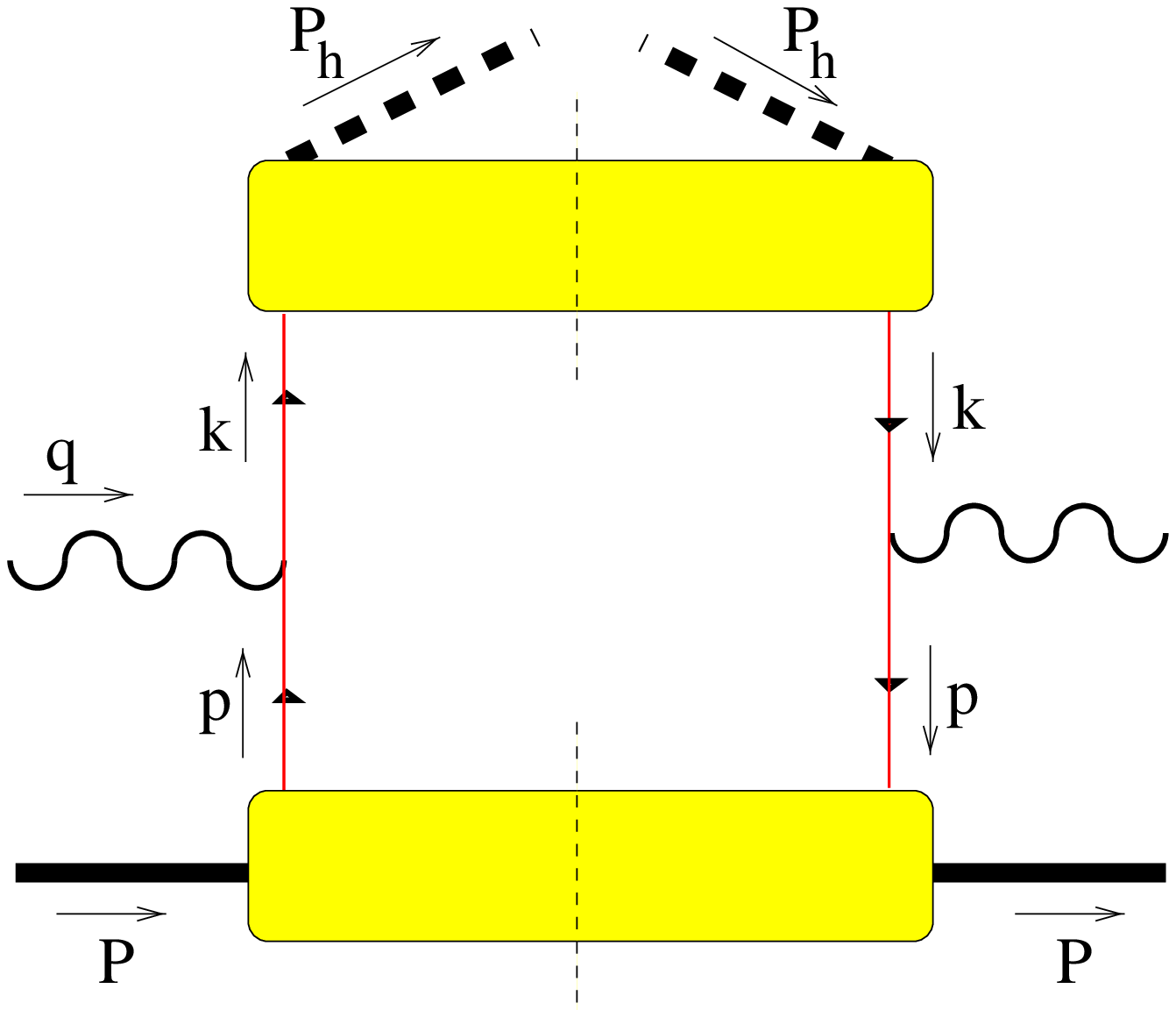}
\end{center}
\caption{\label{fig2}\em
The simplest (parton-level) diagrams representing the squared amplitude
in lepton hadron inclusive scattering (left) en semi-inclusive scattering
(right).}
\end{figure}

In order to calculate the hadronic tensor, a diagrammatic expansion is
written down starting with the well-known handbag diagram
(see Fig.~\ref{fig2},left),
yielding the parton model results for the structure functions, 
\bea
&&F_T(\xbj,Q) = F_1(\xbj,Q)
= \frac{1}{2}\sum_{a,\bar a} e_a^2\,f_1^a(\xbj),
\\ && F_L(\xbj,Q) = 0,
\eea
expressed in terms of the quark distribution $f_1^a$ ($a$ is the flavor index).
The summation runs over quarks and antiquarks.
The most general antisymmetric part of the hadronic tensor involves polarized
leptons and hadrons and is for $\gamma$-exchange given by
\be
2MW_A^{\mu\nu}(q,P,S) =
\underbrace{-i\,\lambda\,
\frac{\epsilon^{\mu\nu\rho\sigma}P_\rho q_\sigma}
{P\cdot q}}_{-i\,\lambda\,\epsilon_\perp^{\mu\nu}}
\,g_1
+ i\,\frac{2M\xbj}{Q}
\,\hat t_{\mbox{}}^{\,[\mu}\epsilon_\perp^{\nu ]\rho} S_{\perp\rho}
\,g_T
\ee
with $\lambda \equiv q\cdot S/q\cdot P$ and $S_\perp$ the transverse spin
vector obtained with the help of $g_\perp^{\mu\nu}$. The cross section
becomes
\be
\frac{d\sigma_L}{d\xbj dy} = \lambda_e\,\frac{4\pi\,\alpha^2}{Q^2}
\Biggl\{ \lambda\,\left( 1-\frac{y}{2} \right) g_1
-\vert S_\perp\vert\,\cos\,\phi_S^\ell\,\frac{2M\xbj}{Q} \sqrt{1-y}
\,\,g_T\Biggr\},
\ee
with the parton model results
\bea
&&g_1(\xbj,Q)
= \frac{1}{2}\sum_{a,\bar a} e_a^2\,g_1^a(\xbj),
\\ && g_T(\xbj,Q) =
(g_1+g_2)(\xbj,Q) =
\frac{1}{2}\sum_{a,\bar a} e_a^2\,g_T^a(\xbj).
\eea
The function $g_1^a$ is the quark helicity distribution. The function
$g_T^a$ is a higher twist distribution.

Proceeding to the 1-particle inclusive case for unpolarized lepton and 
hadron\footnote{
\begin{eqnarray*}
&&\hat q^\mu = q^\mu/Q, \quad
\hat t^\mu = (q^\mu + 2\xbj\,P^\mu)/Q, \quad
\\ &&
q_T^\mu = q^\mu +\xbj\,P^\mu - P_{h}^\mu/\zh =
-P_{h\perp}^\mu/\zh \equiv -Q_T\,\hat h^\mu.
\end{eqnarray*}
}
we obtain generally for the symmetric part of the hadronic tensor
\bea
2M{\cal W}_S^{\mu\nu}(q,P,P_h) &=&
- g_\perp^{\mu\nu}\,{\cal H}_T
\nonumber
+ \hat t^\mu\hat t^\nu\,{\cal H}_L
\\ &&
+ \hat t^{\,\{\mu}\hat h^{\nu\}}\,{\cal H}_{LT}
+ \left\lgroup 2\,\hat h^\mu \hat h^\nu + g_\perp^{\mu\nu}\right\rgroup
{\cal H}_{TT},
\eea
leading to the unpolarized cross section
\bea
&&\frac{d\sigma_O}{d\xbj dy\,d\zh d^2q_T}
= \frac{4\pi\,\alpha^2\,s}{Q^4}\,\xbj \zh
\Biggl\{ \left( 1-y+\frac{1}{2}\,y^2 \right) {\cal H}_T
+ (1-y)\,{\cal H}_L
\nonumber
\\ && \hspace{1.7 cm}
- (2-y)\sqrt{1-y}\,\,\cos \phi_h^\ell\,\,{\cal H}_{LT}
+ (1-y)\,\cos 2\phi_h^\ell\,\,{\cal H}_{TT}
\Biggr\}.
\eea
We will come back to the parton expressions for these structure functions 
later with emphasis on the azimuthal dependence, the $\cos \phi_h^\ell$
and $\cos 2\phi_h^\ell$ parts depending on the azimuthal angle between
the lepton scattering plane and the production plane (see Fig.~\ref{fig1}).
Limiting ourselves to unpolarized leptons, the antisymmetric part of the
hadronic tensor is
\be
2M{\cal W}_A^{\mu\nu}(q,P,P_h) =
- i\hat t^{\,[\mu}\hat h^{\nu]}\,{\cal H}^\prime_{LT},
\ee
leading to the cross section
\be
\frac{d\sigma_L}{d\xbj dy\,d\zh d^2q_T}
= \lambda_e\,\frac{4\pi\,\alpha^2}{Q^2}\,\zh
\,\sqrt{1-y}\,\sin \phi_h^\ell\,\,{\cal H}^\prime_{LT}.
\ee
Our aim in studying leptoproduction is the study of the
quark and gluon structure of the hadronic target using the known
framework of Quantum chromodynamics (QCD). Thus, as a theorist the aim is
to calculate the hadronic tensor $W_{\mu\nu}$ by making a diagrammatic
expansion. Already at the simplest level (Fig.~\ref{fig2}) a problem 
is encountered,
namely there are hadrons involved for which QCD does not provide rules. Thus,
{\em soft parts} are identified that allow inclusion of hadrons in the
field theoretical framework. Furthermore it will turn out that for $Q^2
\rightarrow \infty$ only a limited number of diagrams is needed.

\section{Soft parts}

\subsection{Definition as quark operators}

Next, we look in more detail to the soft parts, such as appear
for instance in the parton diagram. They can be written down in 
terms of quark and gluon fields
as illustrated below. They are characterized by the fact that
the momenta are {\em soft} with respect to each other.
We have for the distribution part~\cite{Soper77,Jaffe83}
\vspace{0.1 cm}\newline
\begin{minipage}{5.5 cm}
\leavevmode
\epsfxsize=5 cm \epsfbox{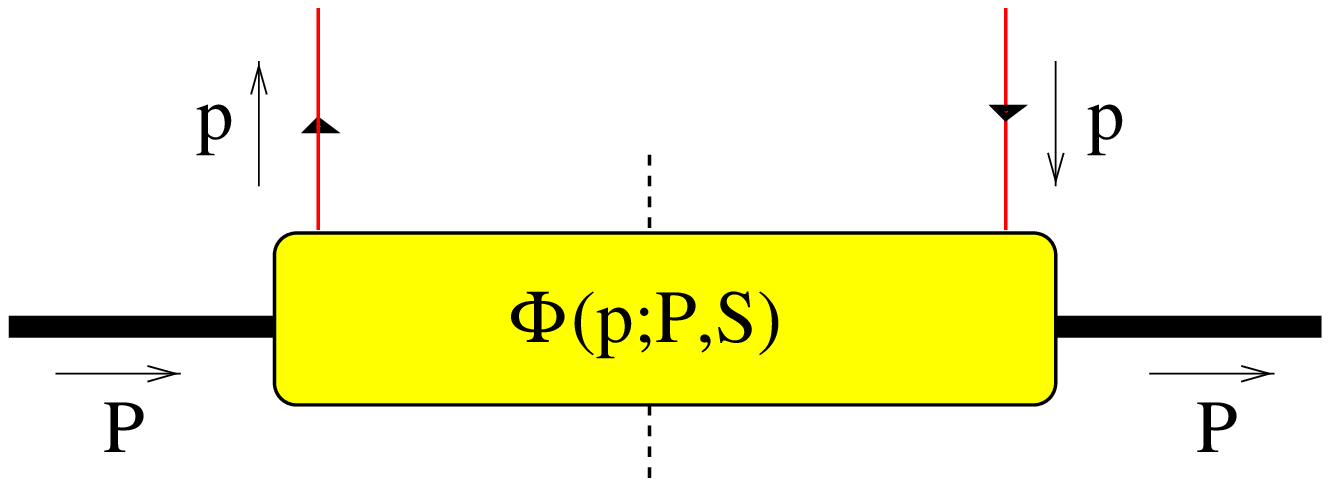}
\end{minipage}
\begin{minipage}{6.5 cm}
\[
\mbox{with}\ \ p^2 \sim p\cdot P \sim P^2 = M^2 \ll Q^2
\]
\end{minipage}
\newline
represented by
\be
\Phi_{ij}(p,P,S)=\frac{1}{(2\pi)^4}\int d^4x \;
e^{ip\cdot x}\;\langle P,S|\overline\psi_j(0)\psi_i(x)|P,S\rangle,
\ee
and the fragmentation part~\cite{CS82}
\vspace{0.1 cm}\newline
\begin{minipage}{5.0 cm}
\leavevmode
\epsfxsize=4.0 cm \epsfbox{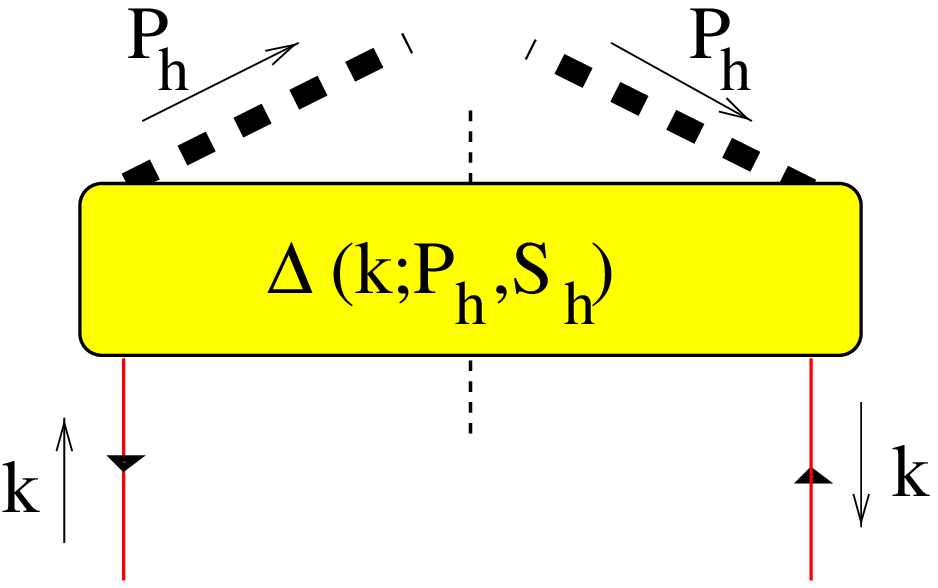}
\end{minipage}
\begin{minipage}{6.5 cm}
\[
\mbox{with}\ \ k^2 \sim k\cdot P_h \sim P_h^2 = M_h^2 \ll Q^2
\]
\end{minipage}
\newline
represented by
\be
\Delta_{ij}(k,P_h,S_h)
=\sum_X\frac{1}{(2\pi)^4}\int d^4x\;
e^{ik\cdot x} \langle 0|\psi_i(x)|P_h,S_h;X\rangle
\langle P_h,S_h;X|\overline\psi_j(0)|0\rangle.
\ee
In order to find out which information in the soft parts $\Phi$
and $\Delta$ is important in a hard process one needs to realize
that the hard scale $Q$ leads in a natural way to the use of lightlike
vectors $n_+$ and $n_-$ satisfying $n_+^2 = n_-^2 = 0$ and $n_+\cdot n_-$
= 1. 
For 1-particle inclusive scattering one parametrizes the momenta
\[
\left.
\begin{array}{l} q^2 = -Q^2 \\
P^2 = M^2\\
P_h^2 = M_h^2 \\
2\,P\cdot q = \frac{Q^2}{\xbj} \\
2\,P_h\cdot q = -z_h\,Q^2
\end{array} \right\}
\longleftrightarrow \left\{
\begin{array}{l}
P_h = \frac{z_h\,Q}{\sqrt{2}}\,n_-
+ \frac{M_h^2}{z_h\,Q\sqrt{2}}\,n_+
\\ \mbox{} \\
q =\frac{Q}{\sqrt{2}}\,n_- - \frac{Q}{\sqrt{2}}\,n_+ + q_T
\\ \mbox{} \\
P = \frac{\xbj M^2}{Q\sqrt{2}}\,n_-
+ \frac{Q}{\xbj \sqrt{2}}\,n_+
\end{array}
\right.
\]
Comparing the power of $Q$ with which the momenta in the soft and hard
part appear one immediately is led to 
$\int dp^-\,\Phi(p,P,S)$ and
$\int dk^+\,\Delta(k,P_h,S_h)$
as the relevant quantities to investigate.
\vspace{0.1 cm}\newline
\begin{minipage}{4.5 cm}
\epsfxsize=4.5 cm \epsfbox{mulders2.eps}
\end{minipage}
\hspace{0.5 cm}
\begin{minipage}{6 cm}
\begin{tabular}{c|cc|c}
part & \multicolumn{2}{c}{'components'} & \\
& $-$ &  + &  \\
\hline
$q\rightarrow h$ & $\sim Q$  &  $\sim 1/Q$  &
$\rightarrow \int dk^+ \ldots$ \\
HARD &  $\sim Q$  &   $\sim Q$  & \\
$H\rightarrow q$ & $\sim 1/Q$  &  $\sim Q$  &
$\rightarrow \int dp^- \ldots$
\end{tabular}
\end{minipage}

\subsection{Analysis of soft parts: distribution and fragmentation functions}
 
Hermiticity, parity and time reversal invariance (T) constrain the quantity
$\Phi(p,P,S)$ and therefore also the Dirac projections
$\Phi^{[\Gamma]}$ defined as
\begin{eqnarray}
\Phi^{[\Gamma]}(x,\bm p_T) & = &
\int dp^- \,\frac{Tr[\Phi \Gamma]}{2}
\nonumber \\ & = &
\left. \int \frac{d\xi^-d^2\bm \xi_T}{2\,(2\pi)^3}\ e^{ip\cdot \xi}
\,\langle P,S\vert \overline \psi(0) \Gamma \psi(\xi)
\vert P,S\rangle \right|_{\xi^+ = 0},
\end{eqnarray}
which is a lightfront ($\xi^+$ = 0) correlation function.
The relevant projections in $\Phi$ that are important in leading order in
$1/Q$ in hard processes are
\begin{eqnarray}
\Phi^{[\gamma^+]}(x,\bpt) & = &
f_1(x ,\bpt^2)
\\
\Phi^{[\gamma^+ \gamma_5]}(x,\bpt) & = &
\lambda\,g_{1L}(x ,\bpt^2)
+ \frac{(\bpt\cdot\bSt)}{M}\,g_{1T}(x ,\bpt^2)
\\
\Phi^{[ i \sigma^{i+} \gamma_5]}(x,\bpt) & = &
S_T^i\,h_1(x ,\bpt^2)
+ \frac{\lambda\,p_T^i}{M} \,h_{1L}^\perp(x ,\bpt^2)
\nonumber \\ && {}
- \frac{\left(p_T^i p_T^j + \frac{1}{2}\bpt^2g_T^{ij}\right)
S_{Tj}}{M^2}\,h_{1T}^\perp(x ,\bpt^2)
\end{eqnarray}
Here $x = p^+/P^+$, $\lambda = MS^+/P^+$ and $S_T$ is the spin-component
projected out by $g_T^{\mu\nu}$ = $g^{\mu\nu} - n_+^{\{\mu}n_-^{\nu\}}$.
They satisfy $\lambda^2 + \bm S_T^2$ = 0.

\begin{figure}[b]
\leavevmode
\epsfxsize=1.8 cm \epsfbox{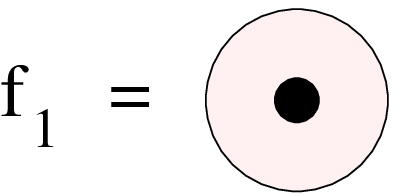}
\newline
\epsfxsize=4.2 cm \epsfbox{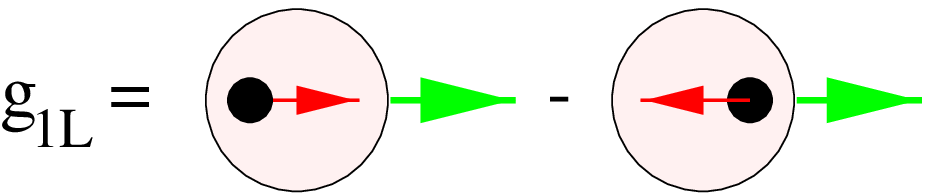}
\hspace{0.5 cm}
\epsfxsize=3.1 cm \epsfbox{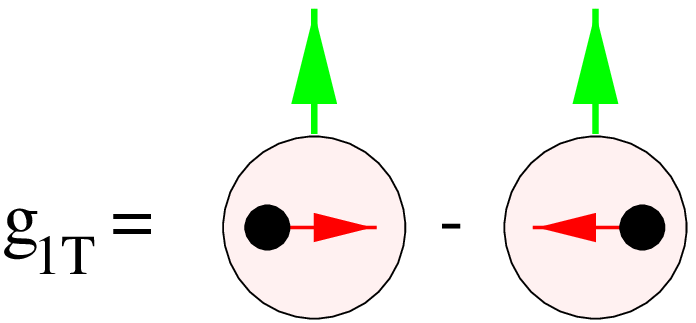}
\newline
\epsfxsize=3.1 cm \epsfbox{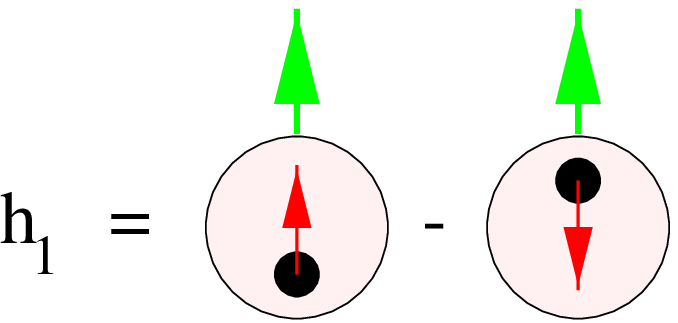}
\hspace{0.5 cm}
\epsfxsize=4.2 cm \epsfbox{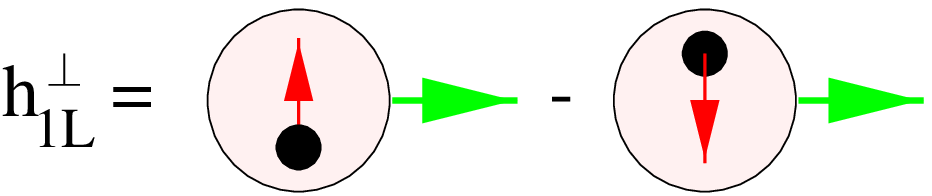}
\hspace{0.5 cm}
\epsfxsize=3.1 cm \epsfbox{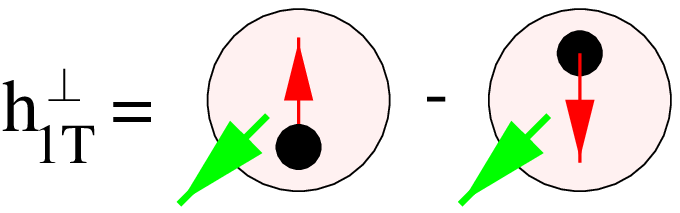}
\caption{\label{fig3}\em
Interpretation of the functions in the leading
Dirac projections of $\Phi$.}
\end{figure}
All functions appearing above can be interpreted as momentum space densities,
as illustrated in Fig.~\ref{fig3}.
The ones denoted~\cite{JJ92} $f_{\ldots}^{\ldots}$
involve the operator structure
$\overline \psi \gamma^+ \psi = \psi_+^\dagger \psi_+$,
where $\psi_+ = P_+\psi$ with $P_+ = \gamma^-\gamma^+/2$. This operator
projects on the socalled good component of the Dirac field, which can be
considered as a {\em free} dynamical degree of freedom in front form
quantization. It is precisely in this sense that partons measured in
hard processes are free. The functions $g_{\ldots}^{\ldots}$ and
$h_{\ldots}^{\ldots}$ appearing above are differences of densities
involving good fields, but in addition projection operators
$P_{R/L} = (1\pm \gamma_5)/2$ and
$P_{\uparrow/\downarrow} = (1\pm \gamma^1\gamma_5)/2$, all of which
commute with $P_+$. To be precise for the functions $g_{\ldots}^{\ldots}$
one has
$\psi \gamma^+\gamma_5 \psi =
\psi_{+R}^\dagger \psi_{+R} - \psi_{+L}^\dagger \psi_{+L}$
while in the case of $h_{\ldots}^{\ldots}$ one has
$\psi \sigma^{1+}\gamma_5 \psi =
\psi_{+\uparrow}^\dagger \psi_{+\uparrow}
- \psi_{+\downarrow}^\dagger \psi_{+\downarrow}$.

It is useful to remark here that flavor indices have been omitted, i.e.
one has $f_1^u$, $f_1^d$, etc. At this point it may also be good to mention
other notations used frequently such as $f_1^u(x) = u(x)$, $g_1^u(x) =
\Delta u(x)$, $h_1^u(x) = \Delta_T u(x)$, etc. These $x$-dependent 
functions are the ones obtained after integration over $\bpt$.

The analysis of the soft part $\Phi$ can be extended to other Dirac
projections. Limiting ourselves to $\bpt$-averaged functions one finds
\bea
& & \Phi^{[1]}(x) =
\frac{M}{P^+}\,e(x),
\\ & & \Phi^{[ \gamma^i \gamma_5]}(x) =
\frac{M\,S_T^i}{P^+} \, g_{T}(x),
\\ & & \Phi^{[ i\sigma^{+-}\gamma_5 ]}(x) =
\frac{M}{P^+}\,\lambda\,h_{L}(x).
\eea
Lorentz covariance requires for these projections on the right hand side
a factor $M/P^+$, which as can be seen from the earlier given parametrization
of momenta produces a suppression factor $M/Q$ and thus these functions
appear at subleading order in cross sections. 
The constraints on $\Phi$ lead to 
relations between the above higher twist functions
and $\bpt/M$-weighted functions~\cite{BKL84,TM96}, e.g.
\be
g_2 \ =\  g_T - g_1 \ =\  \frac{d}{dx}\,g_{1T}^{(1)},
\ee
where 
\be
g_{1T}^{(1)}(x) =
\int d^2\bpt\,\frac{\bpt^2}{2M^2}\,g_{1T}(x ,\bpt) .
\ee
We will use the index $(1)$ to indicate a $\bpt^2$-moment
of the above type.

\begin{figure}[t]
\leavevmode
\begin{center}
\epsfxsize=1.8 cm \epsfbox{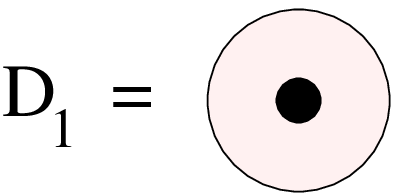}
\hspace{0.5 cm}
\epsfxsize=3.1 cm \epsfbox{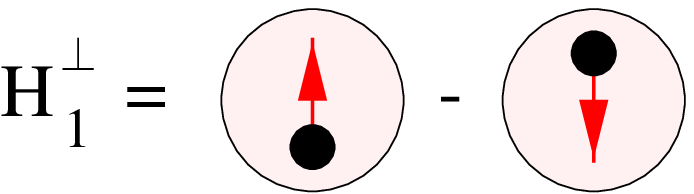}
\end{center}
\caption{\label{fig4}\em
Interpretating the leading 
Dirac projections of $\Delta$ for unpolarized hadrons.}
\end{figure}

Just as for the distribution functions one can perform an analysis of
the soft part describing the quark fragmentation~\cite{TM96}.
The Dirac projections are
\bea
\Delta^{[\Gamma]}(z,\bkt) & = &
\int dk^+\,\frac{Tr[\Delta\Gamma]}{4z}
\nonumber \\ & = &
\left. \sum_X \int \frac{d\xi^+d^2\bm \xi_T}{4z\,(2\pi)^3} \,
e^{ik\cdot \xi} \,Tr  \langle 0 \vert \psi (x) \vert P_h,X\rangle
\langle P_h,X\vert\overline \psi(0)\Gamma \vert 0 \rangle
\right|_{\xi^- = 0}.
\eea
The relevant projections in $\Delta$ that appear in leading order in
$1/Q$ in hard processes are for the case of no final state polarization,
\bea
& & \Delta^{[\gamma^-]}(z,\bkt) =
D_1(z,-z\bkt),
\\ & & \Delta^{[i \sigma^{i-} \gamma_5]}(z,\bkt) =
\frac{\epsilon_T^{ij} k_{T j}}{M_h}\,H_1^\perp(z,-z\bkt).
\qquad \mbox{[T-odd]}
\eea
The arguments of the fragmentation functions $D_1$ and $H_1^\perp$ are
chosen to be $z$ = $P_h^-/k^-$ and $\bm P_{h\perp}$ = $-z\bkt$. The first
is the (lightcone) momentum fraction of the produced hadron, the second
is the transverse momentum of the produced hadron with respect to the quark.
The fragmentation function $D_1$ is the equivalent of the distribution
function $f_1$. It can be interpreted as the probability of finding a
hadron $h$ in a quark. 
Noteworthy is the
appearance of the function $H_1^\perp$, interpretable as the different
production probability of unpolarized hadrons from a transversely
polarized quark (see Fig.~\ref{fig4}). This functions has no equivalent in the
distribution functions and is allowed because of the non-applicability of
time reversal invariance because of the  appearance of out-states
$\vert P_h, X\rangle$ in $\Delta$, rather than the plane
wave states in $\Phi$.

After $\bkt$-averaging one is left with the functions
$D_1(z)$ and the 
$\bkt/M$-weighted result 
$H_1^{\perp (1)}(z)$.
We summarize the full analysis of the soft part with a table of 
distribution and
fragmentation functions for unpolarized (U), longitudinally polarized (L)
and transversely polarized (T) targets, distinguishing leading (twist
two) and subleading (twist three, appearing at order $1/Q$) functions and
furthermore distinguishing the chirality~\cite{JJ92}. 
The functions printed in boldface survive after integration over transverse
momenta. We have for the distributions included a separate table with
distribution functions that can exist without the T constraint,
suggested to explain single spin 
asymmetries~\cite{Sivers90,Anselmino95,Anselmino97}.
We have included them in our complete classification scheme.
\vspace{0.2 cm}\newline
{\bf Classification of distribution and fragmentation functions:}
\begin{center}
\begin{minipage}{6.0 cm}
\begin{tabular}{|c|r|c|c|} \hline
\multicolumn{4}{|c|}{DISTRIBUTIONS (T-even)} \\ \hline
\multicolumn{2}{|c|}{} & \multicolumn{2}{c|}{chirality} \\
\multicolumn{2}{|c|}{$\Phi^{[\Gamma]}$} & even & odd \\ \hline
&  {U} & ${\bm f}_1$ & \\
twist 2 & {L} & ${\bm g}_{1L}$ & $h_{1L}^\perp$ \\
&  {T} & $g_{1T}$ & ${\bm h}_1\ \ {h_{1T}^\perp}$  \\ \hline
&  {U} & ${f^\perp}$ & ${\bm e}$ \\
twist 3 & {L} & ${g_L^\perp}$ & ${\bm h}_L$ \\
&  {T} & ${\bm g}_T\ \ {g_T^\perp}$ &
${h_T}\ \ {h_T^\perp}$  \\
\hline
\end{tabular}
\end{minipage}
\begin{minipage}{5.0 cm}
\begin{tabular}{|c|r|c|c|} \hline
\multicolumn{4}{|c|}{DISTRIBUTIONS (T-odd)} \\ \hline
\multicolumn{2}{|c|}{} & \multicolumn{2}{c|}{chirality} \\
\multicolumn{2}{|l|}{$\Phi^{[\Gamma]}(x,\bkt)$} & even & odd \\ \hline
&  U & $-$ & $h_{1}^\perp$\\
twist 2 & L & $-$ & $-$ \\
&  T & $f_{1T}^\perp$ & $-$
\\ \hline
&  U & $-$ & $\bm h$ \\
twist 3 & L & $f_L^\perp$ & $\bm e_L$ \\
&  T & $\bm f_T$ & $e_T$  \\
\hline
\end{tabular}
\end{minipage}
\end{center}
\begin{center}
\begin{minipage}{9.0 cm}
\begin{tabular}{|c|r|c|c|} \hline
\multicolumn{4}{|c|}{FRAGMENTATION} \\ \hline
\multicolumn{2}{|c|}{} & \multicolumn{2}{c|}{chirality} \\
\multicolumn{2}{|c|}{$\Delta^{[\Gamma]}$} & even & odd \\ \hline
&  {U} & ${\bm D}_1$ & ${H_{1}^\perp}$\\
twist 2 & {L} & ${\bm G}_{1L}$ & ${H_{1L}^\perp}$ \\
&  {T} & ${G_{1T}}\ \ {D_{1T}^\perp}$ & ${\bm H}_1\ \ {H_{1T}^\perp}$ \\
\hline &  {U} & ${D^\perp}$ & ${\bm E}\ \ {\bm H}$ \\
twist 3 & {L} & ${G_{L}^\perp}\ \ {D_L^\perp}\ \ {\bm E}_L$ & ${\bm H}_L$ \\
&  {T} & ${\bm G}_T\ \ {G_T^\perp}\ \ {\bm D}_T\ \ {E_T}$ &
${H_T}\ \ {H_T^\perp}$  \\
\hline \end{tabular}
\end{minipage}
\end{center}

\section{Cross sections for lepton-hadron scattering}

Having completed the analysis of the soft parts, the next step is to find
out how one obtains the information on the various correlation functions
from experiments, in this paper in particular lepton-hadron scattering
via one-photon exchange as discussed in section 1.
To get the leading order result for semi-inclusive scattering it is
sufficient to compute the diagram in Fig.~\ref{fig2} (right)
by using QCD and QED Feynman rules in the hard part and the
matrix elements $\Phi$ and $\Delta$ for the soft parts, parametrized in
terms of distribution and fragmentation functions. The results are:
\newline\newline
\fbox{
\begin{minipage}{11.4 cm}
{\bf Cross sections (leading in $1/Q$)}
\bea
&&\frac{d\sigma_{OO}}{d\xbj\,dy\,dz_h}
= \frac{2\pi \alpha^2\,s}{Q^4}\,\sum_{a,\bar a} e_a^2
\left\lgroup 1 + (1-y)^2\right\rgroup \xbj {f^a_1}(\xbj)\,{ D^a_1}(z_h)
\\ && \frac{d\sigma_{LL}}{d\xbj\,dy\,dz_h}
= \frac{2\pi \alpha^2\,s}{Q^4}\,{\lambda_e\,\lambda}
\,\sum_{a,\bar a} e_a^2\  y (2-y)\  \xbj {g^a_1}(\xbj)\,{D^a_1}(z_h)
\eea
\end{minipage}
}
\newline\newline
Comparing with the expressions in section 1, one can identify
the structure function ${\cal H}_T$ and deduce that in leading order
$\alpha_s^0$ the function ${\cal H}_L$ = 0.

It is not difficult to give some general rules on how the distribution
and fragmentation functions are encountered in experiments~\cite{M96}.
We will just give a few examples.

In 1-particle inclusive processes, one actually becomes sensitive to
quark transverse momentum dependent distribution functions. One finds
at order $1/Q$ the following nonvanishing azimuthal asymmetries~\cite{LM94}:
\newline\newline
\fbox{
\begin{minipage}{11.4 cm}
{\bf Azimuthal asymmetries for unpolarized targets (higher twist)}
\bea
&&\int d^2\bm q_{T}\,\frac{Q_{T}}{M} \,\cos(\phi_h^\ell)
\,\frac{d\sigma_{{OO}}}{d\xbj\,dy\,dz_h\,d^2\bm q_{T}}
= -\frac{2\pi \alpha^2\,s}{Q^4} \,2(2-y)\sqrt{1-y}
\nonumber \\ &&
\hspace{3 cm} \mbox{}\times \sum_{a,\bar a} e_a^2
\Biggl\{
\frac{2M}{Q}\,\xbj^2 {f^{\perp(1)a}}(\xbj)\,{D_1^a}(z_h)
\nonumber \\ && \hspace{4.5 cm} \mbox{}
+\frac{2M_h}{Q}\,\xbj {f_1^a}(\xbj)
\,\frac{{\tilde D^{\perp(1)a}}(z_h)}{z_h}
\Biggr\}
\\ &&
\mbox{note:} \ {\tilde D^{\perp a}}(z) = D^{\perp a}(z) - zD_1^a(z),
\nonumber
\eea
\end{minipage}
}
\newline
\fbox{
\begin{minipage}{11.4 cm}
\bea
&&\int d^2\bm q_{T}\,\frac{Q_{T}}
{M} \,\sin(\phi_h^\ell)
\,\frac{d\Delta\sigma_{{LO}}}{d\xbj\,dy\,dz_h\,d^2\bm q_{T}}
= \frac{2\pi \alpha^2\,s}{Q^4}\,{\lambda_e}
\,2y\sqrt{1-y}
\nonumber \\ &&
\hspace{4 cm} \mbox{}\times \sum_{a,\bar a} e_a^2
\,\frac{2M}{Q}\,\xbj^2 {\tilde e^a}(\xbj)\,{H_1^{\perp (1)a}}(z_h)
\\ &&
\mbox{note:}
\ {\tilde e^a}(x) = e^a(x) - \frac{m_a}{M}\,\frac{f_1^a(x)}{x}.
\nonumber
\eea
\end{minipage}
}
\newline\newline
The first weighted cross section given here involves the structure 
function ${\cal H}_{LT}$ and contains
the twist three distribution function $f^\perp$ and the
fragmentation function $D^\perp$.
The second cross section involves the structure function
containing the distribution function $e$ and the
time-reversal odd fragmentation function $H_1^\perp$.
The tilde functions that appear in the cross sections are in fact
precisely the socalled interaction dependent parts of the twist three
functions. They would vanish in any naive parton model calculation in
which cross sections are obtained by folding electron-parton cross
sections with parton densities. Considering the relation for $\tilde e$
one can state it as $x\,e(x)$ = $(m/M)\,f_1(x)$ in the absence of
quark-quark-gluon correlations. The inclusion of the latter also
requires diagrams dressed with gluons.
\vspace{0.1 cm}\newline\fbox{
\begin{minipage}{11.4 cm}
{\bf Azimuthal asymmetries for unpolarized targets (leading twist)}
\bea
&&
\int d^2\bm q_{T}\,\frac{Q_{T}^2}{MM_h} \,\cos(2\phi_h^\ell)
\,\frac{d\sigma_{{LT}}}{d\xbj\,dy\,dz_h\,d^2\bm q_{T}}
\nonumber \\ && \hspace{2 cm}
= \frac{4\pi \alpha^2\,s}{Q^4}
\,4(1-y)\sum_{a,\bar a} e_a^2
\,\xbj\,{h_{1}^{\perp(1)a}}(\xbj) H_1^{\perp (1)a}.
\eea
\end{minipage}
}

\section{Concluding remarks}

In the previous section some results for
1-particle inclusive lepton-hadron scattering have been presented. 
Several other effects are important in these cross sections, such as
target fragmentation, the inclusion of gluons in the calculation to
obtain color-gauge invariant definitions of the correlation functions and an
electromagnetically gauge invariant result at order $1/Q$ and finally
QCD corrections which can be moved back and forth between hard
and soft parts, leading to the scale dependence of the soft parts and
the DGLAP equations.

In my talk I have tried to indicate why semi-inclusive,
in particular 1-particle inclusive
lepton-hadron scattering, can be important. 
The goal is the study of the quark
and gluon structure of hadrons, emphasizing the
dependence on transverse momenta of quarks. The reason why this prospect is
promising is the existence of a field theoretical framework that allows
a clean study involving well-defined hadronic matrix elements.

\section*{Acknowledgments}
This work is part of the scientific program of the
foundation for Fundamental Research on Matter (FOM),
the Dutch Organization for Scientific Research (NWO)
and the TMR program ERB FMRX-CT96-0008.


\section*{References}
\begin{thebibliography}{99}
\bibitem{Soper77}
D.E. Soper, 
Phys. Rev. {D 15} (1977) 1141; 
Phys. Rev. Lett.  43 (1979) 1847.
\bibitem{Jaffe83}
R.L. Jaffe, 
Nucl. Phys. {B 229} (1983) 205.
\bibitem{CS82}
J.C. Collins and D.E. Soper, 
Nucl. Phys. {B 194} (1982) 445.
\bibitem{JJ92}
R.L. Jaffe and X. Ji, 
Nucl. Phys. {B 375} (1992) 527.
\bibitem{BKL84}
A.P. Bukhvostov, E.A. Kuraev and L.N. Lipatov, 
Sov. Phys. {JETP 60} (1984) 22.
\bibitem{TM96}
R.D. Tangerman and P.J. Mulders, 
\Journal{\NPB}{461}{197}{1996}
\bibitem{Sivers90}
D. Sivers,
\Journal{\PRD}{41}{83}{1990} and
\Journal{\PRD}{43}{261}{1991}.
\bibitem{Anselmino95}
M. Anselmino, M. Boglione and F. Murgia,
\Journal{\PLB}{362}{164}{1995}.
\bibitem{Anselmino97}
M. Anselmino, A. Drago and F. Murgia, hep-ph/9703303.
\bibitem{M96}
P.J. Mulders, in Proceedings of the 2. ELFE Workshop, St. Malo, 23-27 Sept.
1996, nucl-th/9611040.
\bibitem{LM94}
J. Levelt and P.J. Mulders, 
Phys. Rev. {D 49} (1994) 96;
Phys. Lett. {B 338} (1994) 357.
\end{thebibliography}
\end{document}